\documentclass[conference]{IEEEtran}

\usepackage[dvips]{graphicx}
\usepackage[tight]{subfigure}
\usepackage{amsmath}
\usepackage{amssymb}
\usepackage{psfrag}

\newtheorem{theorem}{Theorem}

\begin{document}

\title{On Superposition Coding\\ for the Wyner-Ziv Problem}

\author{
\IEEEauthorblockN{Lorenzo Cappellari}
\IEEEauthorblockA{Dept. of Information Engineering, University of Padova, Italy\\
Email: lorenzo.cappellari@dei.unipd.it}
}

\maketitle

\begin{abstract}
In problems of lossy source/noisy channel coding with side information, the theoretical bounds are achieved using ``good'' source/channel codes that can be partitioned into ``good'' channel/source codes. A scheme that achieves optimality in channel coding with side information at the encoder using independent channel and source codes was outlined in previous works. In practice, the original problem is transformed into a multiple-access problem in which the superposition of the two independent codes can be decoded using successive interference cancellation. Inspired by this work, we analyze the superposition approach for source coding with side information at the decoder. We present a random coding analysis that shows achievability of the Wyner-Ziv bound. Then, we discuss some issues related to the practical implementation of this method.

\end{abstract}

\section{Introduction}
The problem of source coding in presence of some correlated side information at the decoder which is not accessible by the encoder has many applications. In particular, it arises in all sensor networks, where the power for communication is severely constrained \cite{xiong_DSCSensors}, and in sequential source coding of correlated data when for some reasons (i.e.~encoding complexity or robustness against transmission errors) data must be encoded separately, as in video coding \cite{girod_DVC}.

In the lossless case, the minimum achievable transmission rate was found by Slepian and Wolf, who showed that there is no rate penalty with respect to the case in which the side information is also available at the encoder \cite{slepian_noiselessCodingCorrSrc}. In the lossy case, where the goal is to minimize the transmission rate under a given distortion constraint on the reconstructed source, this \emph{rate-distortion function} was found by Wyner and Ziv \cite{wyner_RDFunctSrcCodingSideInfo}. In the latter case, in general, a rate penalty may occur in having the side information only at the decoder. In both cases, the achievability of these rate bounds was shown using randomly generated codes whose codewords are randomly grouped into bins.

In practice, codes having some structure must be used. In the Slepian-Wolf problem (lossless/near-lossless case) all ``good'' practical codes currently employed in channel coding (such as turbo \cite{berrou_turboJCOM} and LDPC \cite{mackay_ldpc} codes) induce a partition of the set of all source outcomes that is in fact a very good binning. Hence, the utilization of these codes leads to practical schemes with performances close to the theoretical bounds \cite{garcia_turbolikeSPM}. Similar results may be also obtained in lossless \emph{multiterminal} network coding \cite{stankovic06_OnCodeDesign}. In the Wyner-Ziv problem (lossy coding) practical code design is more difficult. Zamir et al.~showed that \emph{nested linear codes} (in the binary case) or \emph{nested lattice codes} (in the continuous case) achieve asymptotically the rate-distortion bound \cite{zamir_nestedCodesStructBin}. Similarly, the same structures are shown to achieve asymptotically the \emph{capacity-cost function} in the channel coding problem where side information regarding the channel state is non-causally available at the encoder but not at the decoder \cite{gelfand80_CodingRandomParameters}. But, in practice, \emph{finite-dimensional} linear/lattice codes must be used as proposed in \cite{servetto_LatticeQuantizationWith} or \cite{liu_SlepianWolfCoded} that limit the achievable performances. However, in \cite{liu_SlepianWolfCoded} a scheme is proposed where finite-dimensional \emph{nested quantization} on nested lattices is followed by a second Slepian-Wolf ``binning'' stage in order to limit the performance loss. At high rates, it is shown that the scheme performs as conventional entropy-coded lattice quantization with side information known as well at the encoder. Similar solutions can be employed also in general \emph{multiterminal} source code design \cite{yang08_OnMultiterminalSource}.

For the case of channel coding with additive noise and an additive \emph{interference} signal non-causally known at the encoder but not at the decoder (i.e.~the so-called \emph{dirty paper} problem \cite{costa83_WritingOnDirty}), a practical scheme has been recently proposed \cite{bennatan_SuperpositionCodingFor} which achieves the capacity-cost function using two \emph{independent} rather than nested codes. In particular, the original problem is transformed into an equivalent \emph{multiple-access} problem in which the \emph{superposition} of the two codes is decoded using successive interference cancellation. Very good performances are shown for both the binary \cite{caire_CodingSchemesFor} and the Gaussian \cite{bennatan_SuperpositionCodingSITA} setting. Since the codes can be independently designed, each one of them can be specifically tailored for the purpose it serves. In fact, one of them mainly serves as source code (on which we must be able to perform \emph{quantization}), while the other serves as channel code (on which we must apply some channel-decoding operations with performance close to the one of \emph{joint-typicality} used in random analysis).

Since the problem of lossy source coding with side information is ``dual'' to the problem of noisy channel coding with side information, it may seem straightforward that a similar superposition approach with independent codes may be effectively used in the Wyner-Ziv setting. However, the encoder/decoder of one problem \emph{functionally} corresponds to the decoder/encoder of the other one only under certain hypothesis \cite{pradhan_dualitySrcChnCodSideInfo}. For example, in the binary setting, the two problems are not exactly duals in the sense discussed in \cite{pradhan_dualitySrcChnCodSideInfo}. Moreover, functional duality assumes that both encoder and decoder operate an exact joint-typicality operation, which in practice is not the case. For example, belief-propagation in traditional channel decoding of turbo/LDPC codes roughly corresponds to joint-typicality only if the input to the decoder is ``close'' to an actual codeword, but not in general. Hence, in this paper, we aim to analyze the superposition approach for the Wyner-Ziv problem without relying on duality.

The rest of this paper is organized as follows. In Section \ref{s:review} we review the superposition coding approach for the dirty paper problem. In Section \ref{s:contrib} we analyze the performance of superposition coding for source coding with side information from a random coding perspective. The result of this analysis is then particularized for the binary and the Gaussian case. The issues involved in a practical implementation of this method are discussed in Section \ref{s:issues}. Section \ref{s:concl} summarizes our conclusions.

\section{Superposition Coding for Writing on Dirty Paper}\label{s:review}
Consider the additive memoryless channel
\begin{equation}
Y = X + S + Z \;,\nonumber
\end{equation}
where $X$ is the input to the channel, $S$ is an interference signal known (non-causally) to the encoder and independent from $X$, $Z$ is an unknown channel noise independent from $X$ and $S$, and $Y$ is the channel output. Assume that the channel is used $n$ times, without feedback; $X^n$, $S^n$, $Z^n$, and $Y^n$ denote the involved random vectors.

In the binary case the alphabet over which the random variables take values is $\mathcal{A}=GF(2)$, ``$+$'' indicates the sum over the field $GF(2)$, $S\sim\mathcal{B}(1/2)$ (i.e.~is distributed as Bernoulli-$1/2$), $Z\sim\mathcal{B}(p)$, and $X$ is subject to the cost constraint $E[d_H(X^n,0^n)/n] \leq W$, where $d_H(\cdot,\cdot)$ is the Hamming distance. Two codes $\mathcal{C}_0\subset\mathcal{A}^n$ and $\mathcal{C}_1\subset\mathcal{A}^n$ are constructed at rates $R_0$ and $R_1$ (bit/symbol) by random i.i.d.~selection according to distributions $\mathcal{B}(1/2)$ and $\mathcal{B}(q)$, respectively.

Given $x^n$ and a code $\mathcal{C}$, $T_\mathcal{C}(x^n)$ denotes a codeword of $\mathcal{C}$ which is (strongly) jointly typical with $x^n$ if there exist at least one, otherwise it is a random codeword of $\mathcal{C}$. Define $[x^n]_\mathcal{C} \triangleq T_\mathcal{C}(x^n) - x^n$. The encoder selects a codeword $c_1^n\in\mathcal{C}_1$ and sends the sequence
\begin{equation}
x^n = [s^n - c_1^n]_{\mathcal{C}_0} \;.\nonumber
\end{equation}
If $T_{\mathcal{C}_0}(s^n - c_1^n)\triangleq c_0^n \in\mathcal{C}_0$, the channel output is
\begin{equation}
y^n = c_0^n - (s^n - c_1^n) + s^n + z^n = c_0^n + c_1^n + z^n \;.\nonumber
\end{equation}
The decoder computes the pair $(\hat{c}_0^n,\hat{c}_1^n)\in\mathcal{C}_0\times\mathcal{C}_1$ such that it is jointly typical with $y^n$, announcing $\hat{c}_1^n$ as the decoded message.

By random analysis over all possible tuples of codes (see \cite{bennatan_SuperpositionCodingFor}), if $R_0>R_{1/2}(W)$, where $R_{1/2}(W) = 1 - H(W)$ is the Hamming-distortion rate-distortion function \cite{cover_ElemInfoThNew} of a binary symmetric source, the average probability of having $X^n$ violate the cost constraint $W$ (encoder error) approaches zero with $n\to\infty$.\footnote{Note that in fact $S^n - C_1^n$ is a binary symmetric source.} In addition, if $R_1 \leq C_p(q)$ and $R_0 + R_1 \leq C_p(1/2)$, where $C_p(q) = H(p \ast q) - H(p)$ ($p \ast q \triangleq p(1-q)+q(1-p)$) is the Hamming-weight cost-capacity function of a binary symmetric channel with error probability $p$, the probability of having $(\hat{c}_0^n,\hat{c}_1^n)\ne(c_0^n,c_1^n)$ (decoder error) vanishes with $n\to\infty$.\footnote{Note that in fact $C_0^n + C_1^n + Z^n$ defines a binary symmetric channel in both the case with the codewords of $\mathcal{C}_1$ as input (and known $C_0^n$) or the case with codewords of $\mathcal{C}_0+\mathcal{C}_1$ as input.} If a $q^\ast$ is chosen such that $p \ast q^\ast = W$, then reliable transmission (i.e.~without encoder nor decoder error) is possible at rates arbitrarily close to $R_1^\ast = H(W) - H(p)$, which equals the cost-capacity function with side information at the encoder of the considered channel, at least for all constraints $1 - 2^{-H(p)}\leq W\leq 1/2$.

In the continuous case the alphabet over which the random variables take values is $\mathcal{A}=[-A/2,A/2)\sim\mathbb{R}/A\mathbb{Z}$ (for some $A>0$), ``$+$'' indicates the modulo-$A$ sum, $Z\sim\mathcal{N}_A(0,P_Z)$ (i.e.~is distributed as $A$-aliased Gaussian variable with zero mean and variance $P_Z$), and $X$ is subject to the cost constraint $E[\|X^n\|^2/n] \leq P_X$, where $\|\cdot\|^2$ is the square-distance norm. The two codes $\mathcal{C}_0\subset\mathcal{A}^n$ and $\mathcal{C}_1\subset\mathcal{A}^n$ are constructed at rates $R_0$ and $R_1$ (bit/symbol) by random i.i.d.~selection according to distributions $\mathcal{U}[-A/2,A/2)$ (i.e.~uniform) and $\mathcal{N}_A(0,Q)$, respectively.

Assume that $D$ is a \emph{dither} signal known to both encoder and decoder, drawn accordingly to $\mathcal{U}[-A/2,A/2)$, and independent from all other variables. The encoder selects a codeword $c_1^n\in\mathcal{C}_1$ and sends the sequence
\begin{equation}
x^n = [\alpha s^n - c_1^n + d^n]_{\mathcal{C}_0} \;.\nonumber
\end{equation}
If $T_{\mathcal{C}_0}(\alpha s^n - c_1^n + d^n)\triangleq c_0^n \in\mathcal{C}_0$, the channel output is
\begin{equation}
y^n = c_0^n - (\alpha s^n - c_1^n + d^n) + s^n + z^n \;.\nonumber
\end{equation}
The decoder first computes
\begin{equation}
\hat{y}^n = \alpha y^n + d^n = c_0^n + c_1^n + \underbrace{[\alpha z^n - (1 - \alpha) x^n]}_{\hat{z}^n} \;,\nonumber
\end{equation}
and then finds the pair $(\hat{c}_0^n,\hat{c}_1^n)\in\mathcal{C}_0\times\mathcal{C}_1$ such that it is jointly typical with $\hat{y}^n$, announcing $\hat{c}_1^n$ as the decoded message. $\alpha = P_X/(P_X + P_Z)$ minimizes the power of the equivalent noise in this virtual multiple-access channel (MAC), producing $P_{\hat{Z}} = \alpha P_Z$.

Again, if in $\mathcal{C}_0$ there are enough codewords, it is possible to have a vanishing probability of encoding errors; in addition, if in $\mathcal{C}_1$ and in $\mathcal{C}_0+\mathcal{C}_1$ there are not too many codewords, it is possible to have a vanishing probability of decoding errors. In particular when $A\to\infty$ (i.e.~when the channel is AWGN), if a $Q^\ast$ is chosen such that $P_{\hat{Z}} + Q^\ast = P_X$, then reliable transmission (i.e.~without encoder nor decoder error) is possible at rates arbitrarily close to $R_1^\ast = (1/2)\log_2(1+P_X/P_Z)$, which equals the cost-capacity function with side information at the encoder of the considered channel \cite{costa83_WritingOnDirty}. Details are given in \cite{bennatan_SuperpositionCodingFor}.

Note that in both cases the two codes take a very different role. The code $\mathcal{C}_1$ must be designed in order to be a good channel code with respect to the virtual MAC arising in the decoding operation. The code $\mathcal{C}_0$ has instead dual requirements to be both a good source code, in order to avoid the encoder error, and a good channel code, for effective MAC decoding. However, since we are not interested in exactly decoding the correct $c_0^n$, one should mainly be concerned in choosing a code $\mathcal{C}_0$ which is good for source coding purposes. The same exact considerations arise when analyzing this superposition scheme from a practical perspective. In practice, in source encoding the joint-typicality operation is replaced by maximum likelihood decoding, and structured codes which allow for exhaustive search over all codewords (e.g.~trellis codes) are employed; in channel decoding the joint-typicality operation is approximated using some belief propagation algorithm conducted on codes with a higher degree of randomness (e.g.~turbo or LDPC codes). For these reasons, in practical schemes where $\mathcal{C}_0$ is a trellis code, $\mathcal{C}_1$ is a turbo code, and iterative algorithms are employed during decoding, the performances are very close to the theoretical bounds, even if $\mathcal{C}_0$ is not good from a channel coding perspective.

\section{Superposition Coding for Wyner-Ziv Coding}\label{s:contrib}
The superposition approach for writing on dirty paper is possible because of the additive effect of the interference signal (known at the encoder). Let us consider a tuple of correlated random variables $(X,Y)$ such that
\begin{equation}
X = Y + Z \;,\nonumber
\end{equation}
for some $Z$ independent from $Y$; $X$ is the input to the encoder and $Y$ is the side information (known at the decoder); $\hat{X}$ denotes the reconstruction of $X$ at the decoder. The alphabet over which these random variables take values is a \emph{finite group} $\mathcal{G}$ with $2^l$ elements ($l\geq 1$), ``$+$'' indicates the sum over this group, $Y\sim\mathcal{U}(\mathcal{G})$ (i.e.~is uniformly distributed over $\mathcal{G}$), $Z\sim p(z)$, and $\hat{X}$ is subject to the distortion constraint $E[d_H(X^n,\hat{X}^n)/n] \leq D$.

The two codes $\mathcal{C}_0\subset\mathcal{G}^n$ and $\mathcal{C}_1\subset\mathcal{G}^n$ are constructed at rates $R_0$ and $R_1$ (bit/symbol) by random i.i.d.~selection according to distributions $\mathcal{U}(\mathcal{G})$ and $q(\cdot)$, respectively.

The encoder looks for a codeword $(c_0^n+c_1^n)$, with $c_i^n\in\mathcal{C}_i$, such that it is (strongly) jointly typical with $x^n$, and sends $c_1^n$ to the decoder (if there are no jointly typical codewords, a random codeword is sent). There is an encoder error if $d_H(x^n,c_0^n+c_1^n)>D$.

\begin{theorem}
The probability of having an encoder error vanishes with $n\to\infty$ if
\begin{eqnarray}
R_0 &>& l - H(q \ast d) \nonumber\\
R_0+R_1 &>& R_\mathcal{U}(D) \;,\nonumber
\end{eqnarray}
where $d(\cdot)$ is the distribution which takes zero with probability $1-D$ and all other symbols with probability $\frac{D}{2^l-1}$, and $R_\mathcal{U}(D) = H(X) - H(d)$ is the Hamming-distortion rate-distortion function of the uniform random variable\footnote{$d(\cdot)$ maximizes $H(d)$ and $H(q \ast d)$ over all distributions with $d(0)=1-D$.}.
\end{theorem}

\emph{Proof}: Asymptotically, $x^n$ takes values on a set of $2^{nH(X)}$ elements, and for a fixed $c_0^n$, the code $\mathcal{C}_1$ covers at most $2^{nH(q\ast d)}$ of them within distortion $D$. Hence, there must be at least $2^{nH(X) - nH(q\ast d)}$ codewords in code $\mathcal{C}_0$. Then, since there are at most $2^{nH(d)}$ covered elements within \emph{balls} of distortion $D$, the two codes must provide at least $2^{nH(X)-nH(d)}=2^{nR_\mathcal{U}(D)}$ codewords. Once we have these two conditions the probability of finding a typical $(c_0^n+c_1^n)$ approaches one and the distortion constraint is not violated. This may be proved with an argument similar to the one used in the standard proof of the achievability of the rate-distortion function \cite{cover_ElemInfoThNew}.

The decoder receives the codeword $c_1^n = x^n-c_0^n+\hat{z}^n$, where $\hat{Z}$ is independent from $X$ and distributed as $d(\hat{z})$ (it corresponds to the \emph{subtractive} noise in the equivalent symmetric \emph{test channel} between $\hat{X}$ and $X$). Then it computes
\begin{equation}
[y^n-c_1^n]_{\mathcal{C}_0} = [c_0^n-(z^n+\hat{z}^n)]_{\mathcal{C}_0} \stackrel{c.d.}{=} z^n+\hat{z}^n\label{e:dec}
\end{equation}
and finally reconstructs
\begin{equation}
\hat{x}^n = y^n + [y^n-c_1^n]_{\mathcal{C}_0} \stackrel{c.d.}{=} x^n + \hat{z}^n \;.\nonumber
\end{equation}
Equality in (\ref{e:dec}) is conditional on correct decoding (no decoding errors), i.e.~it holds only if there is only one codeword in $\mathcal{C}_0$ jointly typical with $(y^n-c_1^n)$.

\begin{theorem}
The probability of having a decoder error vanishes with $n\to\infty$ if
\begin{equation}
R_0 \leq C_{p \ast d} \;,\nonumber
\end{equation}
where $C_{p \ast d} = l - H(p \ast d)$ is the (unconstrained) capacity of an additive channel on $\mathcal{G}$ with noise distributed as $p \ast d(\cdot)$.
\end{theorem}

\emph{Proof}: Asymptotically, the equivalent noise $(z^n+\hat{z}^n)$ is distributed according to $p \ast d(\cdot)$ (note that $\hat{Z}$ is independent from $Z$), and hence takes values on a set having $2^{nH(p \ast d)}$ elements. Hence, there can be at most $2^{nH(X)-nH(p \ast d)}=2^{nC_{p \ast d}}$ non overlapping codewords in $\mathcal{C}_0$. Again, once this condition is met the achievability of no decoding errors may be proved with an argument similar to the one used in the standard proof of the achievability of channel capacity \cite{cover_ElemInfoThNew}.

\begin{figure}
\centering
  \begin{small}
    \psfragscanon
    \psfrag{R0}[Br][Bc]{$R_0$}
    \psfrag{R1}[Bc][Bc]{$R_1$}
    \psfrag{Rm}[Bc][Bc]{$R_1^\ast$}
    \psfrag{SB}[Br][Bc]{$R_\mathcal{U}(D)$}
    \psfrag{SB1}[Bc][Bc]{$R_\mathcal{U}(D)$}
    \psfrag{B1}[Br][Bc]{$l - H(q \ast d)$}
    \psfrag{B2}[Br][Bc]{$C_{p \ast d}$}
    \includegraphics[width=0.6\columnwidth]{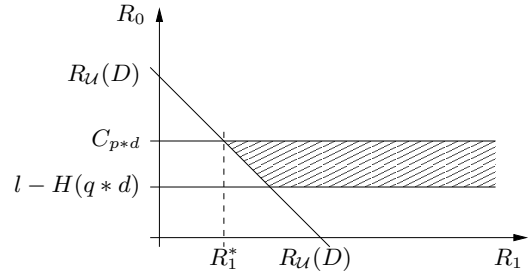}\\
  \end{small}
  \caption{Region in which there are no encoder nor decoder errors.}
  \label{f:region}
\end{figure}

The rate region where there are no errors is not empty for any $q(\cdot)$ such that $H(q\ast d) > H(p\ast d)$ and is shown in Fig.~\ref{f:region}. In the following, we will particularize this result for the doubly symmetric binary and for the Gaussian case, and show that the Wyner-Ziv bound can be achieved in both cases.

\subsection{Binary Sources}
In case of $\mathcal{G}=GF(2)$, where $Z\sim \mathcal{B}(p)$ and $C_1\sim \mathcal{B}(q)$, the rate-distortion function and the channel capacity involved in the calculation of the rate region equal
\begin{eqnarray}
R_\mathcal{U}(D) &=& 1 - H(D) \nonumber\\
C_{p \ast d} &=& 1 - H(p \ast D) \;,\nonumber
\end{eqnarray}
respectively. Hence, the lowest achievable value for $R_1$ is
\begin{equation}
R_1^\ast = R_\mathcal{U}(D) - C_{p \ast d} = H(p \ast D) - H(D) \;,\nonumber
\end{equation}
that is the rate-distortion function with side information for all distortions $0\leq D\leq D'< p$ \cite{wyner_RDFunctSrcCodingSideInfo}. The rate-distortion function for $D'< D\leq p$ is achieved by time-sharing of the two working points $(H(p \ast D') - H(D'),D')$ and $(0,p)$.\footnote{$D'$ is such that $\left.\frac{d[H(p \ast D) - H(D)]}{dD}\right|_{D'} = \frac{H(p \ast D') - H(D')}{D'-p}$.}

\subsection{Gaussian Sources}
Assume now that $\mathcal{G}=\mathbb{R}$, $Y\sim\mathcal{N}(0,P_Y)$, $Z\sim\mathcal{N}(0,P_Z)$, $C_0\sim\mathcal{N}(0,P_0)$, and $C_1\sim\mathcal{N}(0,Q)$. The constraint is given in terms of the squared Euclidean distance. In order to analyze the Gaussian case, which is not discrete, some care should be taken because of the fact that while the \emph{test channel} between $\hat{X}$ and $X$ is still additive, the equivalent channel between $X$ and $\hat{X}$ is additive only if a suitable scaling factor is introduced \cite{cover_ElemInfoThNew}.

In this case, assuming that $U\sim\mathcal{N}(0,D)$ is a \emph{dither} signal known to both encoder and decoder, and independent from all other variables, the encoder sends the $c_1^n$ corresponding to $(\beta x^n+u^n)$, i.e.~the decoder receives
\begin{equation}
c_1^n = \beta x^n+u^n-c_0^n+\hat{z}^n \;,\nonumber
\end{equation}
where $\hat{Z}$ is (a scaled version of) the additive noise in the channel from $X$ to $\hat{X}$, and is independent from $X$ (and from $Z$). If the rate $R_0$ and the sum $R_0+R_1$ are high enough, there exist codes such that $\hat{Z}$ has maximum power $D$ (no encoder error).

The decoder evaluates
\begin{equation}
[\beta y^n+u^n-c_1^n]_{\mathcal{C}_0} = [c_0^n-(\beta z^n+\hat{z}^n)]_{\mathcal{C}_0} \stackrel{c.d.}{=} \beta z^n+\hat{z}^n\label{e:decG}
\end{equation}
and finally reconstructs
\begin{equation}
\hat{x}^n = y^n + \beta[\beta y^n+u^n-c_1^n]_{\mathcal{C}_0} \stackrel{c.d.}{=} x^n + (\beta\hat{z}^n - (1-\beta^2)z^n) \;.\nonumber
\end{equation}
If $\beta=\sqrt{1-D/P_Z}$, the power of $(\beta\hat{z}^n - (1-\beta^2)z^n)$ is minimized and equals exactly $D$; the power of $(\beta z^n+\hat{z}^n)$ equals $P_Z$. Hence, if the rate $R_0$ is less than a certain threshold, we can have correct decoding in (\ref{e:decG}), i.e.~no decoding error. The minimum achievable rate $R_1^\ast$ in this case can be asymptotically computed with a geometric argument: the final goal is to cover each ball related to a codeword of $\mathcal{C}_0$ (which has at least power $P_Z$) with as least as possible balls of power $D$, each one of them related to one codeword of $\mathcal{C}_1$ (for which $Q+D> P_Z$). Finally we obtain
\begin{equation}
R_1^\ast=\frac{1}{2}\log_2\left(\frac{P_Z}{D}\right)\;,\nonumber
\end{equation}
which equals the rate-distortion function with side information in the Gaussian case. The power $P_0$ must be enough in order to have the codewords of $C_0+C_1$ cover all the space in which asymptotically $\beta x^n+u^n$ lies.

\section{Implementation Issues}\label{s:issues}
As clear from the previous section, with superposition coding it is possible to find two independent codes that guarantee the achievability of the Wyner-Ziv bound in the two examined cases. In particular, the code $\mathcal{C}_1$ must offer a good \emph{covering} of the space, i.e.~it must be good for source coding purposes. On the other side, $\mathcal{C}_0$ must take the role of both a good code for source coding and a good code for channel coding. However, even if $\mathcal{C}_0$ was not very good from a source coding perspective, it is reasonable that increasing the \emph{power} ($q$ or $Q$) of $\mathcal{C}_1$ the superposed code is still good for source coding. It is instead crucial that $\mathcal{C}_0$ is a good channel code in order to avoid the decoding errors.

The best source codes available today are represented by the trellis codes \cite{marcellin_TCQ} which offer performance very close to the rate-distortion function. For this codes, given a random realization of the variable to be quantized, it seems crucial that there is the possibility to search for the closest (i.e.~the most likely) codeword by examining all codewords. Hence, $\mathcal{C}_1$ should in general be a good trellis code.

The best channel codes are instead represented by turbo and low-density parity-check codes (LDPC). Those codes have a higher degree of randomness, which prevents (from a computational complexity point of view) conducting the search over all codewords for applying exactly a maximum likelihood approach. However, there exist very good \emph{message passing algorithms} over their \emph{factor-graphs} \cite{kschischang_FactorGraphsAnd}, which almost always converge to the most likely output. Then, $\mathcal{C}_0$ should be one of these codes.

Unfortunately, the message-passing algorithms fail to converge when the input distribution is not unimodal and centered over an actual codeword, as it happens in quantization. In the superposition coding approach, in principle, we should be able to quantize the source outcome over $\mathcal{C}_0$ and successively quantize a \emph{residual} over $\mathcal{C}_1$ (this would be similar to successive interference cancellation used in the MAC case). For the reason mentioned above, the currently available algorithms do not allow to perform the first quantization. An approach in which $\mathcal{C}_0$ is chosen to be a convolutional code was presented in \cite{cappellari_DSCcontinueSyndrome} and achieved a $3\div 4$ dB gap with respect to the Wyner-Ziv bound. That gap is due to the fact that the performance of convolutional codes is somewhat far from the channel capacity.

For practical implementation of the superposition approach, which still remains interesting for the possibility to use \emph{independent} codes, it will be necessary to find good sparse codes and develop good algorithms for lossy quantization over them. One of the first works showing that iterative algorithms may work for quantization too appeared in \cite{martinian03_IterativeQuantization}; that result was obtained by duality with good codes for binary erasure channels. More recently, schemes based on \emph{survey propagation} were adapted from the field of statistical physics in order to do data compression as well \cite{ciliberti05_messagePassing}. Among the sparse codes that are currently under investigation for source coding, there is an high interest in low-density generator matrix codes (LDGM) which are shown to achieve the rate-distortion bound \cite{martinian06_LowDensityCodes,wainwright_SparseGraphCodes}. Nevertheless, general algorithms which allow for practical utilization of these codes have not appeared yet. Other somewhat more practical approaches have appeared in \cite{wainwright05_LossySourceEncoding,gupta07_NonlinearSparseGraph,filler07_BinaryQuantizationBP}.

\section{Conclusion}\label{s:concl}
In this paper we discussed the superposition coding approach for the problem of source coding with side information at the decoder. For the case of a general additive-symmetric discrete \emph{correlation} channel between the (uniform) side information and the source, we derived a rate region for the two independent superposed codes in which the desired distortion bound can be achieved. We showed that in the binary case the Wyner-Ziv bound is achievable, and extended the same result to the Gaussian case with Gaussian side information. Finally, we discussed the implementation issues involved in this scheme, which requires quantization over a code which must be a ``good'' code from a channel coding perspective.

\bibliographystyle{IEEEtran}
\bibliography{IEEEabrv,nonIEEEabrv,refs_books,refs_DSC,refs_TCQ,refs_other,refs_my}

\end{document}